%% file: main.tex
\documentclass[sigconf,10pt]{acmart}
\renewcommand\footnotetextcopyrightpermission[1]{} 
\usepackage{url}
\usepackage{float}

\usepackage{tabularx,colortbl}
\usepackage{algorithm}
\usepackage{wrapfig}
\usepackage{lipsum}
\usepackage{float}

\usepackage{graphicx,graphics}
\usepackage{xcolor}
\usepackage{algpseudocode}
\usepackage{subcaption}
\usepackage{thmtools, thm-restate}
\usepackage{mathrsfs}
\usepackage{tikz}
\usetikzlibrary{positioning}
\usetikzlibrary{decorations.text}
\usetikzlibrary{decorations.pathmorphing}
\usetikzlibrary{arrows,petri, topaths}
\usepackage[inline]{enumitem}

\input{macros}

\acmDOI{10.475/123_4}

\acmISBN{123-4567-24-567/08/06}

\acmConference[SIGMOD'19]{ACM SIGMOD conference}{June 2019}{Amsterdam}
\acmYear{2019}
\copyrightyear{2019}

\acmArticle{4}
\acmPrice{15.00}

\begin{document}

\title{Providing Insights for Queries affected by Failures and Stragglers}

%
%
\author{Bruhathi Sundarmurthy$^*$}
\affiliation{%
	\institution{Google}
}
\email{bruhathi@cs.wisc.edu}
\thanks{$^*$Work done while at UW-Madison}
\author{Harshad Deshmukh$^*$}
\affiliation{%
	\institution{Google}
}
\email{harshad@cs.wisc.edu}
\author{Paraschos Koutris}
\affiliation{%
	\institution{University of Wisconsin - Madison}
}
\email{paris@cs.wisc.edu}
\author{Jeffrey Naughton}
\affiliation{%
	\institution{Google}
}
\email{naughton@google.com}

\input{abstract}

\maketitle

\input{intro}
\input{related}

\input{background}

\input{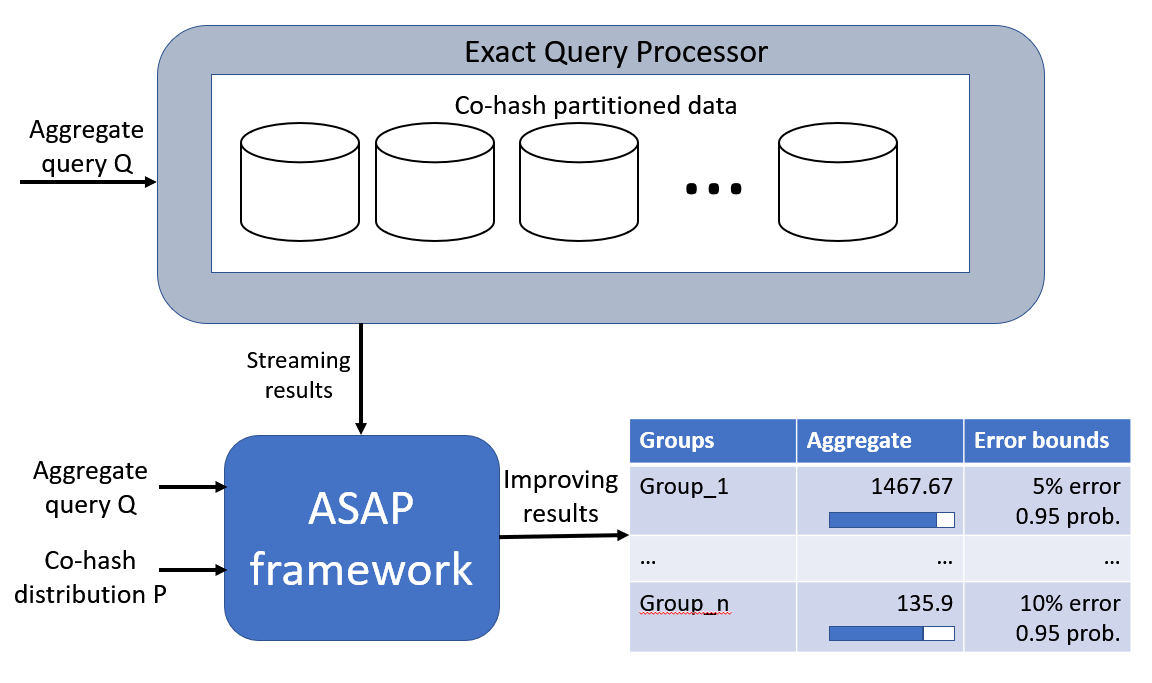}
\input{resultingSample}
\input{heuristic}
\input{queryComplexity}
\input{eval}

\input{conc}

\bibliographystyle{abbrv}
\bibliography{pods}


\end{document}

%% file: macros.tex
\newcounter{reviewcounter}

\newcommand{\inlineitem}[1][]{%
\ifnum\enit@type=\tw@
    {\descriptionlabel{#1}}
  \hspace{\labelsep}%
\else
  \ifnum\enit@type=\z@
       \refstepcounter{\@listctr}\fi
    \quad\@itemlabel\hspace{\labelsep}%
\fi}
\makeatother

\newtheorem{definition}{Definition}
\newtheorem{example}{Example}

\newcommand{\cut}[1]{}

\newenvironment{packed_item}{
\begin{itemize}
   \setlength{\itemsep}{1pt}
   \setlength{\parskip}{0pt}
   \setlength{\parsep}{0pt}
}
{\end{itemize}}

\newenvironment{packed_enum}{
\begin{enumerate}
   \setlength{\itemsep}{1pt}
  \setlength{\parskip}{0pt}
   \setlength{\parsep}{0pt}
}
{\end{enumerate}}

\newcommand{\mS}{\mathcal{S}}

\newcommand{\bU}{\mathbf{U}}
\newcommand{\att}{\mathsf{att}}

\newcommand{\db}{\mathcal{D}}

\newcommand{\pscheme}{\mathcal{H}}
\newcommand{\Var}{\operatorname{Var}}
\newcommand{\E}{\operatorname{E}}

\newcommand{\introparagraph}[1]{\vspace{1mm} \noindent \textbf{#1 }}


%% file: abstract.tex
\begin{abstract}
Interactive time responses are a crucial requirement for users analyzing large amounts of data. Such analytical queries are typically run in a distributed setting, with data being sharded across thousands of nodes for high throughput. However, providing real-time analytics is still a very big challenge; with data distributed across thousands of nodes, the probability that some of the required nodes are unavailable or very slow during query execution is very high and unavailability may result in slow execution or even failures. The sheer magnitude of data and users increase resource contention and this exacerbates the phenomenon of stragglers and node failures during execution. In this paper, we propose a novel solution to alleviate the straggler/failure problem that exploits existing efficient partitioning properties of the data, particularly, \emph{co-hash partitioned} data, and provides approximate answers along with confidence bounds to queries affected by failed/straggler nodes. We consider aggregate queries that involve joins, {\tt group by}s, {\tt having} clauses and a subclass of nested subqueries. Finally, we validate our approach through extensive experiments on the TPC-H dataset. 
\end{abstract}

%% file: intro.tex
\section{Introduction}
\label{sec:intro}


When running queries over data distributed over multiple nodes, nodes fail or are slow with high probability; in such scenarios, systems may return errors or provide degraded performance to the user even though \emph{most of the data} has been processed or is available for processing. In this paper, we propose ideas to continue providing insightful results to users in the event of failures/stragglers, using available data. 

Data warehouses store unprecedented amounts of data and most modern parallel data processing systems extensively use partitioning for high throughput~\cite{snappy, F1, greenplum, shark}. However, the sheer magnitude of the data coupled with the large number of users introduces varied challenges: $(i)$ with data being distributed across a large number of nodes, the probability that some nodes are slow ({\em stragglers}) or unavailable increases drastically. The query either slows down or systems may just fail the query in such scenarios, $(ii)$ resource contention for nodes increases with number of queries and users which further adds to the delay.

%
A prime requirement of data analysts who use such systems is obtaining interactive-time responses for their queries. Studies~\cite{HCI1, HCI2} have shown that a sub-second response time is crucial for their productivity and that failures after queries run for a long time are very frustrating for users. There has been significant research on query progress monitoring~\cite{progress1, progress2, progress3}. However, users still have little control over a failed query execution, despite being able to monitor its progress.

There has been a lot of work on exploring redundancy and fault tolerance techniques to mitigate the effects of failures/stragglers. 
Deploying and maintaining replicas can be very expensive and the overhead of switching over to replicas can be very large~\cite{redundancy}. 
In this work we explore an orthogonal alternate approach: instead of redundancy to reduce the probability of failure, we allow errors and use probabilistic techniques to provide answers despite failures. 



We exploit a key aspect of efficiency in distributed databases, partitioning, and in particular, \emph{co-hash partitioning}, and propose a novel way to exploit the statistical properties of co-hash partitioned data to provide quick approximate answers with confidence bounds in failure/straggler scenarios. 
We consider aggregate queries that involve joins, {\tt{Group By}} and {\tt {Having}} clauses along with a sub-class of nested sub-queries. 

%
We focus on use cases where approximate answers to queries are tolerable, such as gaining insights from data, discovering trends etc.
We note here that there have been no discussion of providing approximate answers for queries in the presence of straggler nodes and node failures. Systems such as Dremel~\cite{dremel} provide an answer based on available data, but without any estimations or attached error bounds. 

{\em Co-hash partitioning}, a well known partitioning strategy, has recently become a popular choice in several systems, and been shown to improve the efficiency of distributed query processing for both OLAP and OLTP workloads~\cite{oracleRef, snappy, F1, jecb,  locationAware}. 
It is an extension of hash-partitioning where tuples from multiple tables that join are co-located by partitioning on join predicates. 
\vspace{-0.07in}
\begin{example}
\label{ex:initial}
Consider the {\tt Customer}, {\tt Orders}, and {\tt Lineitem} tables from the TPC-H schema. Under a co-hash partitioning scheme, the {\tt Customer} table can be hash-partitioned on the {\tt {c\_custkey}} attribute, with all orders and all lineitems of a customer being co-located along with the customer tuple.
\end{example}

%
%
%
%
%
Our key insight is that, with co-hash partitioning, available data from a hierarchy corresponds to obtaining a uniform random \emph{cluster sample}~\cite{cluster} of the hierarchy. 
%
Using this key insight, we propose that a given query be executed to ``completion'' using data that is available or has already been processed and provide an estimate of the query aggregate result along with confidence bounds.

\introparagraph{Users.} They obtain approximate answers with confidence bounds instead of errors due to failures. They are less frustrated, and if the accuracy of the answer is satisfactory, they will not restart the query, thus saving resources and time.

\introparagraph{DB Vendors.} Since databases are co-hash partitioned to aid query performance, our solution requires minimal changes to existing database engines or data layout strategies.
\begin{example}
Consider an aggregate query that joins three tables from the TPC-H benchmark --- {\tt Customer}, {\tt Orders} and {\tt Lineitem}. Continuing Example~\ref{ex:initial}, if all three tables are co-hash partitioned into multiple shards, data in a single shard will itself comprise a {\em cluster sample} of the join result that can be used to provide approximate answers. 
\end{example}
%

\introparagraph{Contributions.}

\introparagraph{Key idea.}  We propose that a widely-employed, efficient partitioning technique, co-hash partitioning, be exploited to provide approximate answers along with confidence bounds quickly in failure/straggler scenarios.


\introparagraph{Feasibility and statistical properties.} For a given query and co-hash partitioning strategy, we enumerate necessary and sufficient conditions to provide approximate answers in failure/straggler scenarios (approximable query). If a query is approximable, we determine the resulting sampling design for the available data. 




\introparagraph{Error bounds.} We discuss providing error bounds for approximate answers. Computing error bounds for queries that involve multiple co-hash hierarchies can be challenging due to correlation between the resulting tuples. We propose a simple heuristic for easing the variance computation; we validate the heuristic experimentally.

\introparagraph{Approximating complex queries.} We formally describe the class of nested sub-queries and queries with {\tt Having} clauses that can be approximated using our proposed idea. 
As an example, consider the {\tt Orders} and {\tt Lineitem} tables being co-hash partitioned. A query such as \emph{``count the number of orders that have at least 10 lineitems"} can be easily approximated, since all lineitems of a order will be co-located. 


\introparagraph{Experimental Validation.} 
We conduct extensive experimentation to validate our proposed approach on uniform and skewed TPC-H datasets using previously validated co-hash partitioning designs~\cite{locationAware}. 

\input{applications}

%% file: applications.tex
\subsection{Other Applications }
\label{subsec:apps}
We now present other applications of the idea proposed:

\introparagraph{Resource Management.} 
Consider a system that has data partitioned across nodes in a network and a large number of users running analytical queries over the partitioned data. Resource contention will be a bottleneck for performance when all users need to access all shards. Since with co-hash partitioning, for many queries, any subset of data can be used provide approximate answers, users can be redirected to disjoint set of machines to significantly increase throughput.
\vspace{-0.2in}
\begin{example}
Consider $m$ users running $k$ instances of query 3 from the TPC-H benchmark that involves a join between {\tt Customer}, {\tt Orders} and {\tt Lineitem}.
Suppose the dataset is co-hash partitioned so that tuples that join are co-located.
If the data is sharded across $M$ nodes, queries from each user can be run on a single unique shard if $M > m$, or the load can be balanced by distributing  $\frac{M}{m \cdot k}$ queries to each shard. This reduces resource contention and increases throughput, with the trade-off of obtaining approximate answers.

\end{example}

\introparagraph{EQP + AQP.}
As shown in~\cite{oracleRef, snappy, jecb} and~\cite{ locationAware}, co-hash partitioning results in huge improvements in performance for EQP (Exact Query Processing). This benefit can carry over to AQP as well; in a nutshell, when data is co-hash partitioned, we can provide approximate answers incrementally by processing queries using efficient exact query processing techniques, without having to repartition data randomly. Also, since co-hash partitioning co-locates joining tuples, it can broadens the scope of AQP to joins and more complex queries that involve nested sub-queries. We argue that since we propose to exploit existing features in EQP systems to provide approximate answers, it may hasten the adoption of AQP in existing systems.
\vspace{-0.07in}
\begin{example}
Consider an analyst who intends to run an aggregate query; she only needs to know a rough estimate of the aggregate but is unsure about the percentage of error that she can tolerate. 
She runs the query using online aggregation. If, after a few minutes, she is not comfortable with the error bounds, it is hard for her to decide whether to wait to obtain better bounds, or whether to stop the query and re-run it using EQP. This is because online aggregation can be inefficient when processing a query all the way to completion.

However, when the data-set is co-hash partitioned, any subset of the database can be used to provide approximate answers, and hence, running estimates of the answer with error bounds can be provided to the user while running the query using EQP. She may stop the query after the desired confidence bounds are reached or may continue the query till the end, with no significant overhead.
\end{example}
\vspace{-0.1in}
%
%
%
%

%% file: related.tex
\section{Related Work}
\label{sec:related}

There is a wealth of relevant work on approximate query processing; we highlight the differences with previous work that is most closely related to ours. To the best of our knowledge, this is the first full-paper that proposes approximate answers with confidence bounds for failure/straggler scenarios. 

\introparagraph{Extended abstract X,} a preliminary publication of this submission (the workshop expects the work to be submitted to a conference later), proposes exploiting partitioning to obtain approximate answers in failure scenarios and otherwise. This current submission extends those ideas in non-trivial ways: (1) X considers only non-redundant designs, whereas, our work considers redundant co-hash partitioning designs as well, (2) We conduct extensive experiments for varied types of queries over multiple recommended co-hash partitioning designs, (3) We discuss in detail, the heuristic for variance computation and validate it empirically, (4) We include a deeper discussion on obtaining estimates and confidence intervals for query affected by stragglers/failures, (5) We propose other applications for the proposed solution, and (6) We extend the ideas to more complex queries with Group By, Having and nested sub-queries.

\introparagraph{Join synopses~\cite{synopses}} results in uniform random samples of the join output. It corresponds to the statistical properties of a subset of a database for a specific co-hash partitioning strategy. However, the samples in~\cite{synopses} are pre-computed, requiring extensive work on top of an existing system and is not designed for handling failures or stragglers.

\introparagraph{Wander Join~\cite{wanderJoin}} provides efficient ways to approximate aggregates in the presence of joins with the help of indexes; however, they do not consider node failures in 
	parallel systems, 
	and the approach can be inefficient if the user chooses to run the query to near completion as Wander Join differs substantially from efficient exact join algorithms. 
	
\introparagraph{BlinkDB~\cite{blink}} handles only single table aggregates. 

\introparagraph{Quickr~\cite{quickr}}: The idea is to obtain samples from different tables from the same hash space. Depending on the co-hash partitioning strategy, the effect can be similar. However, there are fundamental differences -- they aim to push samplers into the query plan to obtain approximate answers and do not discuss handling failures/stragglers.

Systems iOLAP~\cite{iolap} and G-OLA~\cite{gola} provide incrementally improving approximate results. They require an initial random partitioning of data into subsets and do not exploit existing physical data partitioning that many existing systems rely on for efficient performance.  
The idea of randomly partitioning a data-set to aid online aggregation has been suggested and studied in numerous contexts~\cite{onlineAggregation, caseOnlineAgg, parallelHash,
  continuousSampling}. However, in the previous work, the focus is on randomly partitioning each table independently of the other; co-hash partitioning is not considered. 
 Representing all possible answers~\cite{mtables} are not very insightful for aggregate queries and providing partial results~\cite{partial} to users may not be insightful since they neither obtain estimates of the aggregates nor the errors possible in the given result.

%% file: background.tex
\section{Background}
\label{sec:background}

We now provide a brief background on co-hash partitioning and various sampling designs.

\input{partitioning}

\subsection{Sampling Designs and Estimators}
\label{sec:sampling}

A {\em sampling design} refers to an algorithm used to obtain a sample from the input data. 
We introduce the various sampling designs that will be used in the paper in this subsection. The notations are summarized in Table~\ref{tab:notation}.

%
%
%
%
\begin{table}
\small
\begin{tabular}{|l|l|}
\hline
Notation & Meaning \\ \hline \hline
$N$ & number of tuples in table \\
$S$ & current sample \\ \hline
$\pi_i$ & inclusion probability of tuple $i$ \\
$\pi_{ij}$ & inclusion probability of tuples $i$ and $j$\\ \hline
$N_A$ & total number of clusters at level $A = \{I, II \ldots \}$ \\
$C_H$ & clustering attributes at different stages for hierarchy $H$\\
\hline
\end{tabular}
\caption{Notation for Sampling Designs}
 \vspace{-0.1in}
\label{tab:notation}
\end{table}

\subsection{Sampling Designs}
%

\introparagraph{Bernoulli sampling.} In Bernoulli sampling, each element is included in the sample independently with the same probability $p$. The inclusion probabilities are $\pi_i = p$, and $\pi_{ij} = p^2$. The expected size of the sample is $\E[n_S] = N \cdot p$.
%

\introparagraph{Cluster Sampling~\cite{cluster, bookSarndal}.}  In cluster sampling, the table is divided into a number of disjoint groups, called {\em clusters}\footnote{This is not be confused with the dataset being divided into partitions for storing -- each partition can store multiple clusters.}. Then, a sample of the clusters is selected, and all the tuples in the clusters are included in the sample. Cluster sampling is also referred to as single-stage sampling.

\begin{example}
\label{ex:cluster_sample}
Consider the tuple distribution in Table~\ref{tab:cohash} of hierarchy $H_6$. If we obtain a random sample of customers using custkey, then we also get the relevant orders and lineitems of those customers since they are co-located. This constitutes a cluster sample of the data in hierarchy $H_6$.
\end{example}


\introparagraph{Multi-stage Sampling~\cite{multistage, bookSarndal}.}  In multi-stage sampling, we use the subscripts $I, II \ldots$ to refer to the different stages of sampling. 
The tuples in the table are initially partitioned into $N_I$ clusters represented by set $C_I$ and in the first stage, a sample $S_I$ is drawn from $C_I$.
In the second stage, each cluster $i \in C_I$  is further partitioned
into $N_{II}$ secondary clusters represented by set $C_{IIi}$.
For each cluster $i \in S_I$, a sample $S_{IIi}$ is drawn from the $N_{II}$ elements. We repeat this procedure of sampling sub-clusters for $r$
stages. At the $r$-th stage, the sampling unit can be tuples or clusters.  The inclusion probability of cluster $i$ at stage one is $\pi_{Ii}$ and so on. The clustering attributes of the different stages is given by $C_H$, where $H$ the hierarchy that is being sampled. 

\begin{example}
	\label{ex:multi_stage_sample}
Continuing Example~\ref{ex:cluster_sample}, if after sampling customers at random, we also sample each customer's orders, followed by each order's lineitems, then we obtain a 3-stage sample. In this case, the clustering attributes of the three stages are given by the primary keys of the three tables, respectively.
\end{example}


%% file: partitioning.tex
\subsection{Co-Hash Partitioning}
\label{subsec:partitioning}

We first set up some necessary notation.
We assume a relational database $\db$ with relational schema
 $\Sigma = \{ R_1(\bU_1),$ $R_2(\bU_2), \ldots, R_n(\bU_n) \}$. Here, $R_i$ is a relation name, and
 $\bU_i$ is the attribute vector of the relation $R_i$. We denote by $\att(R_i)$ the set of
 attributes of relation $R_i$, and by $\mathsf{key}(R_i)$ the attributes that form the {\em primary key} for $R_i$. Furthermore, we denote by $R^\db$ the instance of relation $R$ in the database $\db$.
 
 \introparagraph{Join Graph.}
Given two tables $R(\mathbf{A}), S(\mathbf{B})$, we define a {\em join
  condition} between $R, S$ to be an expression of the form $R.A_1 =
S.B_1 \land R.A_{2} = S.B_{2} \land \ldots \land R.A_{k} = S.B_{k}$, where
$\{A_1, \dots, A_k\} \subseteq \att(R)$ and $\{B_1, \dots, B_k\} \subseteq \att(S)$.

\tikzset{
	table/.style={
		rectangle,
		rounded corners,
		draw=black, thick,
		minimum height=2em,
		text centered},
	pil/.style={
		shorten <=2pt,
		shorten >=2pt,}
}

\begin{figure*}[!ht]
	\begin{subfigure}{0.45\textwidth}
		\centering
		\resizebox{.82\linewidth}{!}{
			\begin{tikzpicture}[node distance=1.2cm, auto]
			\small
			\node[table] (C) {\texttt{Customer}};
			\node[table, right=of C] (O) {\texttt{Order}};
			\node[table, right=of O] (L) {\texttt{Lineitem}};
			\node[table, below=of L] (P) {\texttt{Part}};
			\node[table, below=of P] (PS) {\texttt{Partsupp}};
			\node[table, below=of O] (S) {\texttt{Supplier}};
			\node[table, below=of C] (N) {\texttt{Nation}};
			\node[table, below=of N] (R) {\texttt{Region}};
			
			\path (C) edge[pil] node {\scriptsize custkey} (O) ;
			\path (C) edge[pil] node {\scriptsize nationkey} (N) ;
			\path (N) edge[pil] node {\scriptsize regionkey} (R) ;
			\path (O) edge[pil] node {\scriptsize orderkey} (L) ;
			\path (N) edge[pil] node {\scriptsize nationkey} (S) ;
			\path (L) edge[pil] node {\scriptsize partkey} (P) ;   
			\path (PS) edge[pil] node {\scriptsize partkey} (P) ; 
			\path (PS) edge[pil] node {\scriptsize suppkey} (S) ;
			\path (S) edge[pil] node {\scriptsize suppkey} (L) ; 
			\path (L) edge[pil,bend left=55] node[text width=0.6cm] {\scriptsize suppkey partkey} (PS) ; 
			\end{tikzpicture}}
		\caption{A schema graph for the TPC-H schema constructed using PK-FK joins between the tables.} \label{fig:tpchJoinGraph}
	\end{subfigure}%
	\begin{subfigure}{0.45\textwidth}
		\centering
		\resizebox{.82\linewidth}{!}{
			\begin{tikzpicture}[node distance=1.2cm, auto]
			
			\node[table, fill=blue!20] (C) {\texttt{Customer}};
			\node[table, right=of C, fill=blue!20] (O) {\texttt{Order}};
			\node[table, right=of O, fill=blue!20] (L) {\texttt{Lineitem}};
			\node[table, below=of L,fill=green!20] (P) {\texttt{Part}};
			\node[table, below=of P,fill=green!20] (PS) {\texttt{Partsupp}};
			\node[table, below=of O, dotted] (S) {\texttt{Supplier}};
			\node[table, below=of C, dotted] (N) {\texttt{Nation}};
			\node[table, below=of N, dotted] (R) {\texttt{Region}};
			
			\path (C) edge[<-,pil,thick] node {\scriptsize custkey} (O) ;
			\path (C) edge[pil,dotted] node {\scriptsize nationkey} (N) ;
			\path (N) edge[pil,dotted] node {\scriptsize regionkey} (R) ;
			\path (O) edge[<-,pil,thick] node {\scriptsize orderkey} (L) ;
			\path (N) edge[pil,dotted] node {\scriptsize nationkey} (S) ;
			\path (L) edge[pil,dotted] node {\scriptsize partkey} (P) ;   
			\path (PS) edge[->,pil,thick] node {\scriptsize partkey} (P) ; 
			\path (PS) edge[pil,dotted] node {\scriptsize suppkey} (S) ;
			\path (S) edge[pil,dotted] node {\scriptsize suppkey} (L) ; 
			\path (L) edge[pil,dotted,bend left=55] node[text width=0.6cm] {\scriptsize suppkey partkey} (PS) ; 
			
			\end{tikzpicture}}
		\caption{SDWithoutRedundancy co-hash partitioning strategy for TPC-H. Tables with the same color belong in the same rooted tree.} \label{fig:hierarchy}
	\end{subfigure}%
	\caption{An example of co-hash partitioning for the TPC-H benchmark.} 
	\label{fig:co-hash}
\end{figure*}
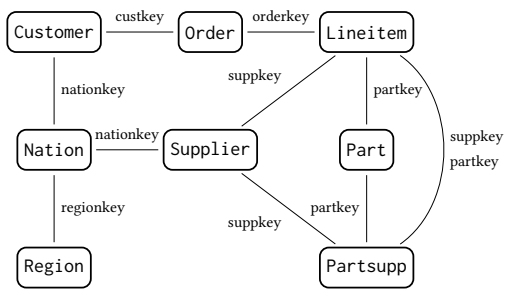
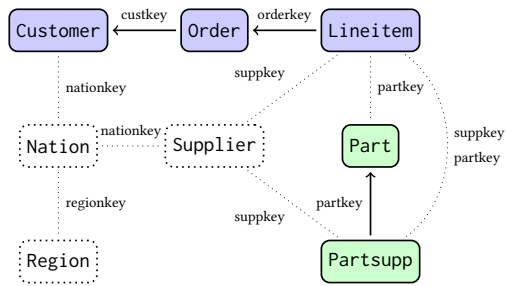

\begin{definition}[Join Graph]
A {\em join graph} is an edge-labeled undirected graph $G = ( V,
E, \lambda) $, where the vertex set $V$ is a subset of $\{R_1, \ldots, R_n\}$, and
the label $\lambda(e)$ of an edge $e = \{R_i, R_j\}$ is a join
condition between $R_i$ and $R_j$.
\end{definition}

Let $Q$ be a Select-Project-Join (SPJ) SQL query.
Then, we can define the join graph $G_Q$ of $Q$ as follows: its vertex set is the
relations involved in $Q$, and there is an edge between two relations whenever these
join in $Q$. 

The {\em schema graph} of a schema $\Sigma$ is a join graph, where the vertex set consists of 
all the relations in $\Sigma$, and each edge represents a possible join between the two tables. 
There are numerous possible schema graphs for a given schema. Figure~\ref{fig:tpchJoinGraph} shows the 
schema graph of the TPC-H benchmark~\cite{tpch} constructed using the primary key-foreign key joins between the tables. The vertices represent the 8 tables in the
schema, and the edges represent the primary key-foreign key joins, as indicated by the edge labels.

\introparagraph{Partitioning Strategies.} 
We now discuss partitioning a database $\db$ into $M$ {\em chunks}.
%
A simple strategy is to partition each table $R_i$ independently of the other tables.  For instance, we can {\em hash-partition} each table $R_i$ using a hash function $h_i$ and a subset of
attributes $\{A_1, \dots, A_k\} \subseteq \att(R_i)$: the hash function takes a value for each attribute 
$A_j$ and
returns a value in $\{1,2, \dots, M\}$. Then, the tuple $t \in R_i^\db$ is assigned to chunk $h_i(t.A_1, \dots, t.A_k)$,
where $t.A_j$ denotes the value of tuple $t$ at attribute $A_j$. 

Hash partitioning can ignore relationships between data across tables.
Suppose we have two tables $R(A,B), S(B,C)$, such that $R.B$ is a foreign key to the primary key $S.B$. If $R$ and $S$ are independently hash-partitioned on $R.A$ and $S.B$, respectively, then 
the join $R \Join_{R.B = S.B} S$ requires data shuffling. However, if we exploit the join predicate between $R$ and $S$ and hash-partition both tables on attribute $B$, then the join can be computed locally without any data shuffling. 
This idea has been previously proposed in multiple systems~\cite{ oracleRef, jecb, locationAware}. Here, we describe formally a more general partitioning framework, called {\em co-hash partitioning}.

\introparagraph{Co-hashing.}  Suppose table $R$ is already distributed
into $M$ (not necessarily disjoint) chunks $R^{(1)}, R^{(2)},
\dots, R^{(M)}$. Let $S$ be a table that shares an edge with $R$ in
the schema graph. Adopting the terminology from~\cite{locationAware}, we
say that $S$ is {\em co-hashed with respect to $R$} if the tuples from
$S$ are distributed as follows:
\begin{packed_item}
\item If $s \in S^\db$ joins with $r \in R^\db$ on the join
  condition $\lambda(\{R,S\})$, then $s$ belongs to all the chunks
  where $r$ belongs; in other words, $s$ is co-located with $r$.
\item If a tuple $s \in S^\db$ does not join with any tuple from $R^\db$ on
  the join condition $\lambda(\{R,S\})$, then $s$ is hash-partitioned
  using any subset of $ \att(S)$.
\end{packed_item}

\introparagraph{Co-hashing Hierarchies.}  We can extend co-hashing
from two tables to a hierarchy.  Let $T$ be a
rooted tree that is a subgraph of the schema graph. We extend co-hashing
on $T$ by following the recursive structure of the tree.  Initially,
we hash-partition the root table $R$ using a subset of its
attributes. We then recursively distribute each table in the tree by
applying co-hashing w.r.t. its parent table. 
Formally, we have:
\begin{definition}[Co-hash Scheme]
Let  $G = (V,E,\lambda)$ be a schema graph of $\Sigma$.
A {\em co-hash hierarchy} $H$ is a tuple $\langle T_H, A_H \rangle$ such that:
\begin{packed_enum}
\item $T_H$ is a rooted directed in-tree~\footnote{An rooted in-tree is a directed tree, where every edge is directed towards a designated root.} that is a subgraph of the schema graph; and
\item $A_H$ is a subset of the attributes $\att(R)$ of the root $R$, 
called the {\em clustering attributes} of $H$.
\end{packed_enum} 
A {\em co-hash scheme} $\mathcal{H}$ is defined as a collection of co-hashing hierarchies
of the schema graph.
%
\end{definition}
Table~\ref{tab:cohashhierarchies} lists some co-hashing hierarchies for the schema graph in Figure~\ref{fig:tpchJoinGraph}. 
If $S$ is co-hashed w.r.t. $R$, then joining $R$ and $S$ on the join
condition $\lambda(\{R,S\})$ (or any superset of the condition) can be
performed without any data shuffling. 

\begin{figure*}
	\begin{subtable}[b]{0.5\textwidth}
		\small
		\centering
		\begin{tabular}[b]{|l|}
			\hline
			$H_3 = {\langle {\tt Customer} \leftarrow {\tt Orders}, \{ {\tt c\_custkey}\} \rangle }$\\
			$H_6 = {\langle {\tt Customer} \leftarrow {\tt Orders} \leftarrow {\tt Lineitem}, \{ {\tt c\_custkey}\} \rangle }$\\
			$H_7 = {\langle {\tt Part} \leftarrow {\tt PartSupp}, \{ {\tt p\_partkey} \} \rangle }$\\
			$H_9 = {\langle {\tt Part} \leftarrow {\tt PartSupp}  \leftarrow {\tt Lineitem}, \{ {\tt p\_partkey} \} \rangle }$\\
			$H_{11} = {\langle {\tt Supplier}, \{ {\tt s\_suppkey} \} \rangle }$ \\
			\hline
		\end{tabular}
		\caption{Example co-hashing hierarchies for the schema graph in Figure~\ref{fig:tpchJoinGraph}. The first element is the hierarchy; the root of the hierarchy is hashed on the attribute(s) given by the second element.} \label{tab:cohashhierarchies}
	\end{subtable}%
	\begin{subfigure}[b]{0.55\textwidth}
		\small
		\begin{subtable}[b]{0.3\textwidth}
			\centering
			\begin{tabular}{|l|l|}
				\hline
				{\tt custkey} & {\scriptsize$\dots$} \\
				\hline
				\color{red} 100 & {\scriptsize\color{red}$\dots$} \\
				\color{red} 101 & {\scriptsize\color{red}$\dots$} \\
				\color{blue} 200 & {\scriptsize\color{red}$\dots$} \\
				\color{blue} 201 & {\scriptsize\color{red}$\dots$} \\
				300 & {\scriptsize$\dots$} \\
				\hline
			\end{tabular}
			\subcaption{{\tt Customer}}
		\end{subtable}%
		\begin{subtable}[b]{0.3\textwidth}
			\centering
			\begin{tabular}{|l|l|l|}
				\hline
				{\tt custkey} & {\tt orderkey} & {\scriptsize$\dots$} \\
				\hline
				\color{red}100 & \color{red}101010 & {\scriptsize\color{red} $\dots$} \\
				\color{red}101 & \color{red}101011 & {\scriptsize\color{red} $\dots$} \\
				\color{blue} 200 & \color{blue} 202020 & {\scriptsize\color{blue} $\dots$} \\
				\color{blue} 200 & \color{blue} 202024 &{\scriptsize\color{blue} $\dots$} \\
				300 & 303030 & {\scriptsize$\dots$} \\
				\hline
			\end{tabular}
			\subcaption{{\tt Orders}}
		\end{subtable}%
		\begin{subtable}[b]{0.5\textwidth}
			\centering
			\begin{tabular}{|l|l|}
				\hline
				{\tt orderkey} & {\scriptsize$\dots$}  \\
				\hline
				\color{red}101010 & {\scriptsize\color{red} $\dots$} \\
				\color{red}101010 & {\scriptsize\color{red} $\dots$} \\
				\color{red}101011 &{\scriptsize \color{red} $\dots$} \\
				\color{blue} 202024 & {\scriptsize\color{blue}$\dots$}  \\
				303030& {\scriptsize$\dots$} \\
				\hline
			\end{tabular}
			\subcaption{{\tt Lineitem}}
		\end{subtable}
		\caption{\small{Tuple distribution for hierarchy $H_6$ from Table~\ref{tab:cohashhierarchies}.}}
		\label{tab:cohash}
	\end{subfigure}
	\caption{Co-hashing hierarchies and an example of their distribution}
\end{figure*}
Multiple co-hash partitioning strategies have been studied in~\cite{locationAware} in terms of performance. They recommend the Schema Driven Without Redundancy (SDWithout) design as the best design without redundancy for the TPC-H benchmark. This design in presented in Figure~\ref{fig:hierarchy} and explained in detail in the example below:
\begin{example}
In the SDWithout design depicted in Figure~\ref{fig:hierarchy}, {\tt Nation}, {\tt Region} and {\tt Supplier} are very small and hence replicated to all partitions (they
are designated with a dotted box).  The remaining five tables are
spanned by two hierarchies, each one depicted with a different
color. The thick arrows show the orientation of the edges in the
hierarchy.
The first hierarchy includes {\tt Customer}, {\tt Order}, {\tt
  Lineitem} and has root the table {\tt Customer} ($H_6$) and the second
hierarchy contains {\tt Part} which is the root and {\tt
  Partsupp} ($H_7$).
Consider hierarchy $H_6$ 
with {\tt custkey} as the hashing attribute for the root table {\tt Customer}.
Figure~\ref{tab:cohash} shows the distribution of tuples in the tables into three different partitions.
\end{example}
In our study we consider both redundant and non-redundant co-hash partitioning designs for providing approximate answers to queries affected by failures/stragglers.

%% file: framework.tex
\section{Preliminaries}
\label{sec:framework}


\introparagraph{Notation.}
Let $\mathcal{D}$ be the database, and $\mathcal{D}^1, \mathcal{D}^2, \dots, \mathcal{D}^M$ denote the $M$ chunks of the database, as partitioned by a co-hash scheme $\pscheme$. For every table $R$ (hierarchy $H$), let $R^i$ ($H^i)$ denote the chunk of the table (hierarchy) in the chunk $\mathcal{D}^i$. A chunk can be thought of as a page, a block (contiguous set of pages) or a node in a cluster, depending on the granularity of the read access. Also, for some chunk $\mathcal{D}^i$, $R^i$ or $H^i$ can be empty; this handles the case where different co-hashed hierarchies are partitioned across disjoint sets of nodes. Suppose that we obtain a subset $\mS \subseteq \{1, \dots, M\}$ of the chunks in the partition. Then, for each table $R$ we obtain a subset $\{ R^i \mid i \in \mS\}$ of its chunks, and for each hierarchy $H$ we obtain also a subset $\{ H^i \mid i \in S\}$ of its chunks. We denote the subset of the database obtained this way by $\mathcal{D_S}$.

The specifics of how these subsets will be obtained in practice is not important; we discuss in Section~\ref{sec:resultingSample} that when a database is co-hash partitioned, the statistical properties of the partitions are agnostic to the way the data is accessed. A hierarchy could be accessed by reading partitions successively or simultaneously. It can also be read via indexes; if table $R$ is accessed through an index using values from attribute $S.A$ of table $S$, then, $R$ and $S$ will be implicitly part of a hierarchy, ($\langle S \leftarrow R, S.A\rangle$), though not explicitly co-hashed. 

\introparagraph{Query class considered.} We consider queries with no self-joins (without repeated occurrences of the same relation) of the form $\mathcal{Q}$:
%
%
\begin{center}
\begin{tabular}{|p{6cm}|}
\hline
Select $A_1, A_2, \ldots, A_{k_1}, agg_1, agg_2, \ldots, agg_{k_2}$ \\
From $R_1, R_2, \ldots, R_n$ \\
Where $(j_1 \land j_2 \land \ldots \land j_l) \land (p_1 \land p_2 \land \ldots \land p_m)$\\
Group By $A_1, A_2, \ldots, A_{k_1}$\\
Having $h_1, h_2, \ldots, h_{k_3}$\\
\hline
\end{tabular}
\end{center}
%
where, $agg_1, agg_2, \ldots, agg_{k_2}$ are one of SUM(), COUNT(), and AVG() aggregate operators, $j_1, j_2, \ldots, j_l$ are equality join conditions, $R_1, \ldots, R_n$ are data sources in or derived from $\mathcal{D}$.
$p_1, p_2, \ldots, p_m$ are predicates that can also be nested {\tt Exist} clauses
and $h_1, h_2, \ldots, h_{k_3}$ are simple predicates.
%
%

\introparagraph{Problem Statement.}
Let's say that while executing a query $Q$ of form $\mathcal{Q}$, some of the nodes required by $Q$ are very slow or unavailable. The problem we consider is as follows:

\begin{center}
\begin{tabular}{|c| p{6.3cm}|}
\hline
Input & Query $Q$ of form $\mathcal{Q}$ (no self-joins), \\
& co-hash scheme $\pscheme$ and \\
& hierarchies chosen to answer the query: $H_Q =\{H_1, \ldots, H_l \}$. \\
\hline
Problem 1 & Can we provide an approximate answer to $Q$ using available data? \\
Problem 2 & Is yes, what are the estimates and confidence intervals?\\
\hline
\end{tabular}\\
\end{center}

We study these questions in detail in the next two sections. Before proceeding, we would like to emphasize that the aim of this work is not to study already established runtime benefits of co-hash partitioning, but rather, to exploit it to provide approximate answers under failures/straggler scenarios. 

%% file: resultingSample.tex
\section{Providing Insights}
\label{sec:resultingSample}

In this section, we first discuss obtaining insights for simple queries of the form {\tt{SELECT <attributes> FROM <tables> WHERE <predicates>}}, without nested sub-queries, {\tt Group By} or {\tt Having} clauses, and then extend the ideas to general queries. 
%
We structure the discussion as follows: (1) failure model and the implied statistical properties of a data-subset (2) necessary conditions that should be satisfied to provide insights for a query in the presence of stragglers and failures, and (4) extensions of the ideas to provide insights to complex queries of the form $\mathcal{Q}$.

\subsection{Statistical properties of data}

We first discuss the implicit statistical properties of data that is co-hash partitioned. This is will help us understand the properties of available data in the event of failures and/or stragglers. We then discuss the failure/straggler model, and the resulting statistical properties of available data.

\introparagraph{Cluster sample}
Consider a hierarchy $H = \langle T_H, A_H \rangle$, where $R_1$ is the root of $H$. From the co-hash partitioning definition, a tuple $r \in R_1$ will be hashed to a machine, and hence will be located in a randomly chosen machine. The set of tuples of $R_1$ located at a machine $M_i$ corresponds to a uniform random sample of $R_1$. 

For a tuple $r_i \in R_i$, let $T_H(r_i)$ denote the set of all tuples in any child $R_j$ of $R_i$ in the co-hash hierarchy $H$, that join with $r_i$ on the join condition $\lambda(R_i, R_{j})$. This is defined as:
$$
	T_H(r_i) = \{\forall R_j \rightarrow R_i \in H, r_{j} \in R_{j} \mid r_i \bowtie r_{j} \text{ on } \lambda(R_i, R_{j})\}
$$
Now, we define the set of tuples from all relations in a co-hash hierarchy $H$ that are co-located with a tuple $r_i \in R_i$, denoted by $\mathcal{J}_H(r_i)$ and defined recursively as follows:
$$
	\mathcal{J}_H(r_i) = T_H(r_i) \cup \bigcup_{r_{j} \in T_H(r_i)} T_H(r_{j}) \cup \ldots
$$
$\mathcal{J}_H(r_i)$ represents the set of all tuples in the co-hash hierarchy $H$ that recursively join with tuple $r_i$. For example, consider a tuple $r_1 \in R_1$, where $R_1$ is the root of the hierarchy $H$. $\mathcal{J}_H(r_i)$ represents the set of all tuples in $R_2$ that join with $r_1$ on $\lambda(R_1, R_2)$, the set of tuples in $R_3$ that join with the preceding set of tuples from $R_2$ and so on.
In other words, all tuples of $R_2, \ldots, R_k$ that recursively join with $r_1$ or some tuple that joins with $r_1$ will be co-located with $r_1$. We call such a set of tuples as a \emph{cluster} with the clustering attribute being the hashing attribute of the root of the hierarchy. That is, $r_1$ along with all its co-located tuples from all relations in the co-hash hierarchy $H$ forms a \emph{cluster} with clustering attributes $A_H$ .

Since $r_1$ is sent to a machine that is randomly chosen, by implication, a cluster of tuples corresponding to $r_1$ are sent to a machine that is randomly chosen. Hence, data in a machine corresponds to a random set of clusters of a co-hash hierarchy.


\begin{example}
	\label{ex:hierarchy}
	Consider hierarchy $H = \langle {\tt Customer} \leftarrow {\tt Orders},$ $\{\tt c\_custkey\} \rangle$. If we look at the data in any machine, it will contain a subset of the clusters of hierarchy $H$. These clusters of a machine will be equal to a cluster sample of the hierarchy with the clustering attribute being $c\_custkey$. 
\end{example}

\introparagraph{Redundancy.} Co-hashing
may result in data redundancy in two ways: (1) a tuple may join with
tuples that belong in different chunks, in which case it will be replicated. This is referred to as \emph{tuple-level redundancy}, (2) a relation can be part of multiple co-hashing hierarchies and this is referred to as \emph{relation-level redundancy}.  

Redundancy has an affect on the statistical properties of the data obtained from a machine and hence, affects the insights we can provide to queries.
When we obtain clusters corresponding to a hierarchy from a machine, it is equivalent to obtaining cluster samples of the result of the join of the relations on the join conditions in the hierarchy. However, the statistical properties of data pertaining to part of the hierarchy or sub-hierarchy depends on whether co-hash partitioning induces tuple-level redundancy for the data corresponding to the relations in the sub-hierarchy considered. 
This is illustrated in the example below:

\begin{example}
	\label{ex:redundant_hierarchy}
	Consider the hierarchy $\langle {\tt Lineitem} \leftarrow {\tt Orders}$, $\{\tt l\_linenumber\}\rangle$ for the join graph in Figure~\ref{fig:tpchJoinGraph}. A tuple $t \in {\tt Orders}$ can join multiple tuples from the table {\tt Lineitem}, and hence, $t$ may be replicated across chunks. A subset of data of only the {\tt Orders} relation will have redundant tuples, and the statistical properties of this data subset is not easy to determine. 
\end{example}
Non-redundancy of a relation can be determined from a given co-hash scheme $\pscheme$ using the following conditions. 
%
\begin{proposition}
	\label{prop:nonredundant}
	A relation $S$ in a co-hashing hierarchy $H$ is {\em non-redundant} if either:
	\begin{packed_enum}
		\item $S$ is the root of $H$; or 
		\item $S$ is co-hashed with respect to $R$ on join condition $\bigwedge_{i=1}^k (A_i = B_i)$ 
		s.t. $ \mathsf{key}(R) \subseteq \bigcup_{i =1}^k \{ B_i \}$, and $R$ is non-redundant. 
		%
		%
	\end{packed_enum} 
\end{proposition}
The proofs of all propositions in this section are simple and were discussed in the extended abstract X as well.

\begin{example}
	\label{ex:nonredundancy}
	Consider the hierarchy $H_3$, also given in Example~\ref{ex:hierarchy}. {\tt Orders} is co-hashed with respect to {\tt Customer} on the join condition {\tt o\_custkey = c\_custkey},  and $\mathsf{key}({\tt Customer}) =$ {\tt c\_custkey}. {\tt Customer} is also the root of $H_3$. Hence, both {\tt Orders} and {\tt Customer} are non-redundant relations.
\end{example}

For a sub-hierarchy with more than one relation, data for the sub-hierarchy will be redundant if the relation corresponding root of the sub-hierarchy is redundant.

\introparagraph{Failure Model.} We assume that the hash functions used for hash partitioning the root tables belong to the universal family of hash functions~\cite{universalHashing} so that the probability of any cluster being mapped to a particular partition is $1/M$, and these probabilities are independent across clusters. Hence, each cluster will be distributed independently and uniformly at random across the $M$ chunks.

For a hierarchy $H$, let $s_H$ denote the number of clusters in the chunk $H^i$. Then, the total number of clusters from $H$ in $\mS$ is $\sum_{i \in \mS} s_H$. If the total number of clusters in hierarchy $H$, $N_I$, is known, then $\mS$ is a cluster sample of $H$, where the inclusion probability of a cluster is $\pi = \sum_{i \in \mS} s_H/ N_I$. Otherwise, we can estimate the total number of clusters as $M \cdot  \sum_{i \in \mS} s_H / |\mS|$.

When the data subset obtained is the result of partitions being unavailable (failure) or slow (a straggler), the resulting inclusion probabilities from the interaction of two random processes -- slow/failed data nodes and co-hash partitioning of tuples/clusters -- may not be immediately apparent. 
%
%

%
\begin{table}
	\small
	\begin{tabular}{|l|l|}
		\hline
		Notation & Meaning \\ \hline
		$t$ & true value of the aggregate \\
		$t_i$ & attribute value at tuple $i$\\ 
		$\hat{t}$ & estimated value of the aggregate \\
		\hline
	\end{tabular}
	\caption{Notation for failure model}
	\vspace{-0.1in}
	\label{tab:notation2}
	\vspace{-0.1in}
\end{table} 
We will now show that the available data after failures can be used as a cluster sample with inclusion probabilities resulting from co-hash partitioning. 
Indeed, let us consider the following failure model for arbitrary elements $i$ and $j$ of a table ($i$ and $j$ could be tuples or clusters). Let $p_f$ be the probability of node failure. (This value need not be known; data nodes in a data center might require monitoring for an extended period of time to determine the expected probability of failure.) Let $\mathcal{S}_1$ be the sample without any failures and let $\mathcal{S}_2 \subseteq \mathcal{S}_1$ be the sub sample set after failures. In the case where $\mathcal{S}_1$ equals the whole table, $\mathcal{S}_2$ represents the sample for which we are deciding the sampling design. The notation for the following equations are given in Tables~\ref{tab:notation} and~\ref{tab:notation2}.
\begin{align}
\label{eq:failureModeleq}
Pr(i \in \mathcal{S}_2 \mid \mathcal{S}_1) =  \pi_i = 1 - p_f 
\end{align}

Now, suppose we define a new estimator for the SUM aggregate (following~\cite{bookSarndal}) as follows:
\begin{align*}
\hat{t} = N \frac{\Sigma_{i \in \mathcal{S}_1} t_i / \pi_i }{\Sigma_{i \in \mathcal{S}_1} 1/\pi_i}
\end{align*}

Adopting the failure model given in Equation~\ref{eq:failureModeleq}, we modify the previous estimator as follows:
\begin{align*}
\hat{t_f} = N \frac{\Sigma_{i \in \mathcal{S}_2} t_i / \pi_i (1 - p_f) }{\Sigma_{i \in \mathcal{S}_2} 1/\pi_i (1 - p_f)}
= N \frac{\Sigma_{i \in \mathcal{S}_2} t_i / \pi_i}{\Sigma_{i \in \mathcal{S}_2} 1/\pi_i}
\end{align*}

Since the $p_f$ factor vanishes, we can easily estimate the aggregate as before -- using inclusion probabilities resulting from co-hashing. 

\introparagraph{Properties of Available Data.}
The core insight from the previous discussion is that the available subset of data for a co-hashed hierarchy resulting from failures and stragglers corresponds to a uniformly random cluster sample (or one-stage sample) of the hierarchy's data, with the clustering attributes being the hashing attributes of the root of the hierarchy. 
 However, since all relations of a hierarchy may not be present in the query, and some relations/sub-hierarchies can be \emph{tuple-redundant}, determining whether a query can be approximated is not trivial. This is illustrated in the following example:
\begin{example}
\label{example:redundant-non-approx}
Consider the hierarchy in Example~\ref{ex:redundant_hierarchy} and consider a query that involves only the {\tt Orders} relation. If all data corresponding to this hierarchy is not available, then, it may not be possible to approximate this query using available data.
\end{example}

\subsection{Feasibility}
%
We now discuss the conditions for determining whether a query can be approximated for a given $\pscheme$.
We are given query $Q$, let $H_Q = \{H_1, \ldots, H_k \} $ be the set of hierarchies used by the query engine to execute $Q$ before the failure or straggler event occurred. For each relation $R$ in the query, there is exactly one  hierarchy $H_R \in H_Q$ from which it's data is being read. Also, if $H_i \in H_Q$, then data of at least one relation present in $H_i$ is being read. The hierarchy from which the data of a relation $R$ is being read is denoted by $H_Q(R)$. Let $H_Q^F$, with $H_Q^F \subseteq H_Q$ be the set of hierarchies for which all data is not available due to failures and/or stragglers. We now discuss determining whether we can provide approximate answers for $Q$ using the available/processed data from $H_Q$. 

We first define a partitioning-query graph (PQ graph). We construct the PQ graph, $G_{PQ}$, as follows. Let $G_Q$ be the join graph of the query $Q$. Then:
\begin{packed_item}
\item For every $v \in V(G_Q)$, we add $v$ to $V(G_{PQ})$.
\item For every $e = (v_i, v_j) \in E(G_Q)$, if there exists a hierarchy $H \in H_Q$ with
$e' = (v_i, v_j) \in H$ and $ \lambda(e) \implies \lambda({e'})$\footnote{Here, the implication means that the join condition $\lambda(e)$ logically implies $\lambda(e')$.}, we add
$e$ to $E(G_{PQ})$.
\end{packed_item}
%

The PQ graph determines the relationship of the tables in $Q$ with respect to the hiearchies they belong to. The connected components in the PQ graph tells us which tables are co-located and hence need to be considered together while determining their statistical properties. Note that the components of a PQ-graph are strictly sub-hierarchies used to answer the query. The root of a component/sub-hierarchy of a PQ-graph can be any table in a hierarchy.

 \begin{example}
\label{ex:PQgraph}
Consider an instance of the TPC-H database partitioned with $\pscheme = \{H_6, H_7, H_{11}\}$, and consider query $Q$: $({\tt Orders} \Join_{{\tt o\_orderkey = l\_orderkey}}$ ${\tt Lineitem})$ $\Join_{\tt{l\_partkey = ps\_{partkey}}} {\tt Partsupp}$. $H_Q = \{H_6, H_7\}$ and $G_{PQ}$ will have two components (or sub-hierarchies): $\{ X_1: {\tt Orders} \leftarrow {\tt Lineitem},  X_2: {\tt Partsupp}\}$.
\end{example}

In order to approximate the result, we require that each connected component of the $PQ$ graph with unavailable data is non-redundant, so that the data subset corresponding to each component/sub-hierarchy forms a cluster sample of the component. For a connected component $X$ of the $PQ$ graph, there is a hierarchy $H_Q$ that corresponds to it. Let $H[X]$ be the hierarchy corresponding to component $X$ and let $R[X]$ be the root of the sub-hierarchy $X$.

\begin{proposition}
\label{prop:nonredundantQ}
For a given $Q$ and $H_Q$, consider $G_{PQ}$. For $X \in G_{PQ}$, if $H[X] \in H_Q^F$ and $X$ is non-redundant, then the available data corresponding to the sub-hierarchy, $X$, is a cluster sample. 
\end{proposition}

\begin{example}
	\label{ex:nonredundant}
Continuing Example~\ref{ex:PQgraph}, if $H_Q^F = \{H_6\}$ and {\tt Orders} ($X_1$) is non-redundant, then, for the available data of $X_1$, we obtain cluster samples. 

However, if $H_Q^F = \{H_6, H_7\}$ and {\tt Orders} ($X_1$) is non-redundant, but {\tt PartSupp} ($X_2$) is not, then we do not obtain cluster samples and we cannot approximate the result of the query.
\end{example}

\begin{definition} [Approximable]
A query $Q$, is said to be approximable for a given $H_Q$ and $H^F_Q$, if, for every connected component $X$ in the  $PQ$ graph $G_{PQ}$, such that $H[X] \in H^F_Q$, the root relation of $X$, $R[X]$ is non-redundant.
\end{definition}


Now, for each $X_i \in G_{PQ}$ with $H[X] \in H^F_Q$, we associate a set of clustering attributes ${B_i}$, sampling probability $\pi_i$, and the sample size (of the available data) $s_{i}$. $B_i$, $\pi_i$ and $s_i$ correspond to the hashing attributes of $H_i$, the available percentage (sampling rate) of $H_i$, and the number of clusters for hierarchy $H_i$ in the data subset, respectively. 
Note that multiple components can correspond to the same hierarchy.

\begin{example}
Continuing Example~\ref{ex:nonredundant} with $H_Q^F = \{H_6\}$, consider $X_1$.  $H_6$ is the corresponding hierarchy of $X_1$.
Let's say we obtain $10000$ tuples out of $20000$ tuples from the {\tt Customer} table; $B_1 = \{ c\_custkey\}$, with $\pi_{1} = 0.5$ and and $s_{1} = 10000$.
\end{example}

%
\input{estimators}

%% file: estimators.tex

\subsection{Estimators and Variance}
\label{subsec:estimators}

Given a query $Q$ that is approximable for a given sets $H_Q$ and $H_Q^F$, we now discuss the resulting sampling design, the corresponding estimators and variance estimators using formulas from~\cite{bookSarndal}.
We start the discussion with the sum aggregate; estimators for count and average aggregates are extensions of it. We first discuss the scenario where data is unavailable for only one hierarchy and then discuss the case when data is unavailable for multiple hierarchies. Notations for the following discussion are listed in Tables~\ref{tab:notation},~\ref{tab:notation2} and~\ref{tab:notation3}.

\introparagraph{Case $|H_Q^F| = 1$.}
When machines that failed or are slow contain data corresponding to only one hierarchy, $H$, i.e, $H_Q^F = \{H\}$, and data is available for all other hierarchies in $H_Q$, available data results in a cluster sample, with the clustering attribute being the attributes corresponding to $R[H]$. 

\begin{table}
	\small
	\begin{tabular}{|l|l|}
		\hline
		Notation & Meaning \\ \hline
		$c_i$ & true value of the aggregate over cluster $i$ \\
		$\hat{c_i}$ & estimated value of aggregate over cluster $i$ \\
		\hline
	\end{tabular}
	\caption{Notation for Estimators}
	\vspace{-0.1in}
	\label{tab:notation3}
\end{table}

The sum aggregate is given by $t = \sum_{i=1}^N c_i$, where $c_i$ is the value of the aggregation for cluster $i$ for hierarchy $H$.
  The simplest way to estimate $t$
from a cluster sample is by using the {\em Horvitz-Thompson (HT) estimator}~\cite{HT},
given by $\hat{t} = \sum_{i \in S} \frac{c_i}{\pi_i}$. 
%
Define $\Delta_{ij} = \pi_{ij} - \pi_i \pi_j$.  The variance and variance estimators for the HT estimator are given by the formula below (adopted from~\cite{bookSarndal}). 
%
\begin{align*}
	\Var(\hat{t}) = \sum_{i = 1}^{N_I} \sum_{j = 1}^{N_I} \Delta_{ij} \frac{c_i}{\pi_i} \frac{c_j}{\pi_j}, \quad
	\widehat{\Var}(\hat{t}) = \sum_{i \in S} \sum_{j \in S} \frac{\Delta_{ij} }{\pi_{ij}}\frac{c_i}{\pi_i} \frac{c_j}{\pi_j} 
\end{align*}

%
The estimator for the count aggregate easily follows from the sum estimator; replace $t_i$, the value of the attribute, with $1$. Estimating the average aggregate involves the ratio of the sum and count estimates, and hence, introduces a slight bias.



\introparagraph{Case $|H_Q^F| > 1$.}  
We now consider the case when machines that are unavailable or slow contain data corresponding to multiple hierarchies. From the previous discussion, we know that the available data for each hierarchy corresponds to a uniform random cluster sample. In the presence of cluster samples from multiple hierarchies, the sampling design of the result of the query involving joins across hierarchies corresponds to a multi-stage sample.

The HT estimator for multi-stage sampling is computed recursively.  Let $c_i$ be the aggregate sum of cluster $i$ from the first stage. Suppose we can compute the HT estimator $\hat{c}_{i}$ for $c_i$ w.r.t. the last $r-1$ stages. Then, the
unbiased HT estimator is given by the formula
$\hat{t} = \sum_{i \in S_I} \frac{\hat{c}_{i}}{\pi_{Ii}} $.

We now discuss the  {\tt SUM} estimator in the context of multi-stage samples obtained through unavailability of data. For a query $Q$, let $\mathcal{X} = \{X_1, \ldots, X_k\}$ be the set of components in $G_{PQ}$ such that, for each $X_i \in X$, $H[X_i] \in H_Q^F$. 
%
The estimator for the $SUM()$ aggregate over is given by
$$
\hat{t} = \frac{\text{Sum over available data}} {\prod_{X_j \in \mathcal{X}} \pi_j}
$$.


\introparagraph{Variance for multi-stage samples.} 
The variance estimate of a multi-stage
sample is broken down into variance contributions from the different
stages. If $V_i$ is the variance of $\hat{c}_{i\pi}$, and $\hat{V}_i$ the
variance estimator, then we can also compute the variance estimate of
r-stage sampling as:
%
%
\begin{align*}
\widehat{\Var}(\hat{t}) = \hat{V}_{stage_1}  + \hat{V}_{stage_2} + \ldots + \hat{V}_{stage_r}
\end{align*}  

%
For 2-stage sampling, the of the second stage ($\hat{c}_{i}$) and the total variance are given below:
\begin{align*}
\hat{V}_i = \sum_{k \in S_i} \sum_{l \in S_i}  \frac{\Delta_{kl|i}}{\pi_{kl|i}}\frac{t_k}{\pi_{k|i}} \frac{t_l}{\pi_{l|i}}\\
\widehat{\Var}(\hat{t}) = \hat{V}_{stage_1}  + \hat{V}_{stage_2}
=  \sum_{i \in S_I} \sum_{j \in S_I} \hat{\Delta}_{Iij} \frac{\hat{c}_{i}}{\pi_{Ii}} \frac{\hat{c}_{j}}{\pi_{Ij}}
+ \sum_{S_I} \frac{\hat{V}_i}{\pi_{Ii}}
\end{align*} 
More details about variance and variance estimation formulas can be found in~\cite{bookSarndal}.
For variance computation, at each stage, the sampling design is assumed to be {\em
independent} and { \em invariant}: invariance implies that the sampling design used a is the same across all stages, and independence implies that the sampled
elements within a cluster are independent of the sampling done in any
other cluster of the same stage.

In the current scenario, the
independence property is violated -- the join involves a cross
product of the samples from different hierarchies and hence, each tuple in the join
result is not independent of the other, and the resulting intermediate table, in
general, will be a correlated sample. There can be correlations in different stages of the multi-stage sample and the independence property requirement may not hold. 

So, the variance estimate can
now be calculated as:
$
\hat{V}(\hat{t}) = \hat{V}_{1} + \hat{V}_{2}^{Corr}  + \ldots + \hat{V}_{k}^{Corr}$, where $ \hat{V}_{i}^{Corr}$ is the correlated variance estimate at the $ith$ stage. Observe that the first stage doesn't have correlation. The formulas and algorithms given in~\cite{largeSampleBounds, crossJoin} to compute the variance and estimated variance apply here as well.

\introparagraph{Cluster sizes and accuracy} 
For single stage (cluster) samples, it is established~\cite{bilevel, bookSarndal} that variance in the sizes of the clusters, and intra-cluster homogeneity (intra-cluster variance is much lower than the inter-cluster variance) leads to higher variance in the estimates. Hence, assuming element (tuple) level randomness instead of cluster level randomness when it is not so, can lead to severe underestimation of variance.  

%% file: heuristic.tex
\section{Efficient variance computation}
\label{sec:heuristic}

Computing variance of multi-stage samples can be cumbersome and compute intensive, especially in the presence of correlations. In a multi-stage sample, the variance contributions decrease with stage, with the first stage contributing majority of the variance. Hence, to make computations easier, the variance can be approximated by the first two terms (the variance contribution from the first two stages of sampling)~\cite{bookSarndal}. This can, however, introduce a slight bias into the variance estimate due to underestimation. 

For the resulting multi-stage samples obtained from available data, we can still apply this approximation, since there will be no correlations in the first-stage samples obtained.  However, note that, this approximation requires that the property of variance contributions of lower stages being larger than the variance contributions of higher stages holds. 
Hence, for computing the variance of a multi-stage sample that is the result of processing available data, mapping the clustering attributes to different stages of the multi-stage sample is very important. This can be non-trivial as illustrated in the following example:

\begin{example}
	\label{ex:design}
	Consider a TPC-H data set that is co-hash partitioned using three hierarchies $H_6$, $H_7$, and $H_{11}$.
	Consider query 9 of the benchmark without the Group By and Order By clauses, whose PQ graph consists of three components, that correspond to the three hierarchies $H_6, H_7,$ and $H_{11}$. Let's say data is unavailable from all three hierarchies. Available data from each hierarchy $H_i$ corresponds to a cluster sample with clustering attributes given by $B_i$. 
	Since there are three sets of clustering attributes, the result obtained after the join, cannot be viewed as a cluster sample of any one set of clustering attributes. It will be a 3-stage sample, with each $B_i$ being the clustering attribute set of one of the stages. However, the mapping between the clustering attributes, $B_i$s, and the different stages so that the majority of the variance contribution is captured in the first two stages is not obvious.
\end{example}

We want to find the clustering attributes that have large contributions to the variance in order to approximate the variance computation. Determining the mapping is crucial, since as we will see later that approximating the variance estimation using an incorrect order underestimates the variance.
We present Algorithm~\ref{algo:mainAlgo} to determine the number of stages in the multi-stage sample for a given $\pscheme$, and the mapping of clustering attributes to the various stages so that the condition that variance contributions of lower stages are larger than the higher stages is satisfied.

\introparagraph{Intuition.} The intuition for the algorithm is: (a) the bigger the clusters, bigger the variance. In fact, sampling from bigger clusters, intuitively make more sense than sampling from smaller clusters, and (b) lower the sampling rate, smaller the number of resulting clusters.

\begin{algorithm}[!tbh]
	\caption{\textsc{Determine Multi-stage Sequence}}\label{algo:mainAlgo}
	\begin{algorithmic}[1]
		\State $\mathcal{X} =  \{X_1, \ldots, X_m\}$; $\forall X_i, H[X_i] \in H_Q^F$
		\State $B_i$: Clustering attribute set of $X_i$
		\State $\pi_{i}$: Sampling probability of $X_i$
		\State $s_{i}$: Number of elements/clusters of $X_i$
		\State Set $stage = 1$;
		
		\While {$\mathcal{X}  \neq \emptyset$}
		\State Find $X_{i} \in \mathcal{X}$ with smallest $s_{i}$.
		
		\For {$X_{j}$ such that $B_{i} \subseteq B_{j}$}
		\State $\pi_i = \pi_i \cdot \pi_j$; $s_{i} = s_{i} \cdot \pi _{j}$;

		$\mathcal{X}  = \mathcal{X}  - X_j$

		\EndFor
		\State Set $B_{i}$ as the clustering attribute set for $stage$ 
		\State Set $\pi_i$ as sampling probability for $stage$
		 \State Set $s_i$ as the cluster size for $stage$
		\State $\mathcal{X}  = \mathcal{X}  - X_i$
		\State $stage$++
		\EndWhile
	\end{algorithmic}
\end{algorithm}

\introparagraph{Working of the algorithm.} It takes as input the list of connected components, their clustering attribute set, the number of clusters, and the inclusion probabilities for each participating component. The algorithm assumes that joining attributes in any two tables have identical names. If this is not the case, it is easy to fix by renaming attributes before performing the join.

Now, starting at the first stage, it determines the clustering attributes for each stage by greedily choosing the component with the minimum number of clusters (line 7). It goes over other connected components and merges them if they have the same set of clustering attributes, and suitably updates the probabilities (lines 8 to 10). 
Finally, it returns the clustering attributes and sampling probabilities for each stage.
We now provide an example to illustrate the working of the algorithm:
\begin{example}
	\label{ex:algoExample}
	Continuing the setup from Example~\ref{ex:design}, we will consider Query 9 and $H_6$, $H_7$ and $H_{11}$ in the co-hash scheme. Let the availability for the hierarchies be 1\% for $H_6$ (rooted at {\tt Customer}), 80\% for $H_7$ (rooted at {\tt Part}), and 90\% for $H_{11}$ ({\tt Supplier}). $G_{PQ}$ consists of 3 components $X_1: Customer \leftarrow Orders \leftarrow Lineitem$, $X_2: Part \leftarrow Partsupp$, and $X_3: Supplier$.  Let $\pi_1 = 0.01, \pi_2 = 0.8$, and $\pi_3 = 0.9$, and $s_1 = 7500, s_2 = 80000$, and $s_3 = 90000$ clusters.
	After renaming so that joining attributes have identical names, 
	we have the component set $\{X_1, X_2, X_3\}$ with $B_1 = \{{\tt c\_custkey}\}, B_2 = \{{\tt
		p\_partkey}\},$ and $B_3 = \{ {\tt s\_suppkey} \}$. 
	$s_3$ corresponding to $X_1$ has the smallest number of clusters (line 7), followed by $s_2$ ($X_1$) and $s_3$
	({$X_2$}). So, Algorithm~\ref{algo:mainAlgo} will determine
	the clustering attributes and inclusion probabilities of the multi-stage sequence to be $(c\_custkey, 0.4)-(p\_partkey, 0.5)-(s\_suppkey, 0.01)$ (c-p-s), in that order. 
\end{example}

\input{accuracy}

%% file: accuracy.tex
\introparagraph{Variance Estimation Accuracy.}
We discuss the impact of sequence on the accuracy of estimated variances for multi-stage samples by comparing different sequences with the sequence given by the algorithm and empirically show that the variance estimation for the multi-stage sequence given by our proposed algorithm is close to the observed variance. 

\begin{figure}
	\centering
	\includegraphics[scale=0.48]{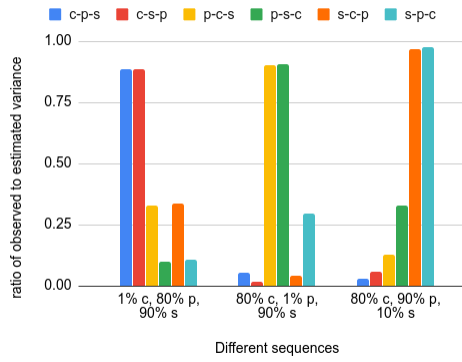}
	\caption{Observed and estimated variance for different sampling rates and multi-stage orders}
	\label{fig:orderMatters}
\end{figure}

We use the query and co-hash partitioning strategy from Example~\ref{ex:algoExample} for the following experiments. For different availability percentages, we plot the ratio of estimated variance and observed variance for different multi-stage sequences in Figure~\ref{fig:orderMatters}. The estimated variance is calculated using uncorrelated variance formulas for the first two stages. There are three hierarchies in the co-hash partitioning graph, $H_6, H_7$ and $H_{11}$ with roots as {\tt Customer}(c), {\tt Part} (p), and {\tt Supplier} (s), respectively. The sequences we consider are c-p-s, c-s-p, p-c-s, p-s-c, s-c-p and s-p-c.

\introparagraph{Combination 1.}
From Example~\ref{ex:algoExample}: Availability percentages are 1\% for $H_6$, 80\% for $H_7$, and
90\% for $H_{11}$ and Algorithm~\ref{algo:mainAlgo} provides the order c-p-s. The ratio of estimated variances to observed variance for all sequences are plotted in the first set of bars in Figure~\ref{fig:orderMatters}.  We see that variance from c-p-s and c-s-p sequences closely match the observed variance, while other sequences severely underestimate the observed variance. This suggests that determining the first stage clustering attributes is crucial.

\introparagraph{Combination 2.}
We now change the availability percentages to be 80\% for $H_{6}$, 1\% for $H_7$, and 90\% for $H_{11}$. Algorithm~\ref{algo:mainAlgo} provides the sequence p-s-c. From Figure~\ref{fig:orderMatters} (middle set of bars), we observe that variance estimated by the p-s-c sequence is closest to the observed variance. 

\introparagraph{Combination 3.}
We set availablity percentages to be 80\% for $H_{6}$, 90\% for $H_7$, and 10\% for $H_{11}$. Algorithm~\ref{algo:mainAlgo} provides the order s-c-p. We observe from the last set of bars in Figure~\ref{fig:orderMatters} that the estimated variance for the sequence s-c-p is very close to the observed variance (ratio of 1).

The above examples validate the insight in Algorithm~\ref{algo:mainAlgo} that the relative number of clusters in hierarchies determine the relative variance contributions of the different stages -- fewer the clusters for a clustering attribute, greater its contribution to the overall variance, with the first stage being very crucial.

%% file: queryComplexity.tex
\section{Extending to General Queries}
\label{subsec:complexity}
We now discuss extending the ideas to include queries containing {\tt Group By}, {\tt Having} clauses and queries with nested sub-queries.

\introparagraph{Group Bys.}  
%
%
Approximating queries with Group By follows the same procedure as basic queries, with estimation and variance estimation being done group-wise. There is a special case that applies whenever cluster sampling (or single stage sampling) is employed -- if the grouping attributes and the clustering attributes are the same, then for any group, we either obtain the exact aggregate values or do not obtain the group. This case occurs if data is unavailable for a single hierarchy, and the clustering attributes of the hierarchy are exactly the same as the grouping attributes of the query.


\introparagraph{Nested query predicates.} A predicate $p$ in the {\tt Where} clause that is a nested query should be a correlated sub-query for the query to be approximable. Let $\mathcal{R}_O$ be the set of relations in the outer query that are correlated with the inner query and let the $\mathcal{R}_I$ be the set of relations in the nested sub query. We will construct the following graph $G_N = \left\langle V, E \right\rangle$ for a given query and co-hash scheme $\pscheme$:
\vspace{-0.08in}
\begin{packed_enum}
\item $\forall R \in \mathcal{R}_O \cup \mathcal{R}_I$, introduce a vertex $v_R \in V$.
\item Let $\lambda'({R_i, R_j})$ be the join conditions between relations $R_i$ and $R_j$ in $Q_p$. If $(R_i, R_j) \in \pscheme$ and $\lambda(R_i, R_j) \subseteq \lambda'(R_i, R_j)$, then add edge $(R_i, R_j)$ to $G_N$.
\end{packed_enum}
After the construction, the number of connected components in $G_N$ should be equal to the number of relations in $\mathcal{R}_O$ in order for the query to be approximable. The intuition behind this condition is that in order for a nested query predicate to be handled, the nested sub-query should only process tuples that are clustered according to the correlated tuples in the outer query.

\introparagraph{Nested query relations.} For multiple input relations, every relation or nested sub-query in the {\tt From} clause should result in a uniform random cluster sample.

\introparagraph{Having clauses.}  We can support {\tt Having} clauses when queries come under the special case of {\tt Group By} clauses.  That is, if we are missing data from a single hierarchy, and the clustering attributes are exactly the same as the grouping attributes of the query, then we obtain complete groups and hence can apply the predicates in the {\tt Having} clauses to provide unbiased estimates and confidence intervals. 

%% file: eval.tex
\section{Evaluation of Co-Hashing}
\label{sec:eval}

\tikzset{
	table/.style={
		rectangle,
		rounded corners,
		draw=black, thick,
		minimum height=2em,
		text centered},
	pil/.style={
		shorten <=2pt,
		shorten >=2pt,}
}

\begin{figure*}
	\begin{subfigure}{0.5\textwidth}
		\centering
		\resizebox{0.82\linewidth}{!}{
			\begin{tikzpicture}[node distance=1.2cm, auto]
			
			\node[table, fill=orange!20] (C) {\texttt{Customer}};
			\node[table, right=of C, fill=orange!20] (O) {\texttt{Order}};
			\node[table, right=of O, fill=orange!20] (L) {\texttt{Lineitem}};
			\node[table, below=of L,fill=orange!20] (P) {\texttt{Partsupp}};
			\node[table, below=of P,fill=orange!20] (PS) {\texttt{Part}};
			\node[table, below=of O,fill=orange!20] (S) {\texttt{Supplier}};
			\node[table, below=of C, dotted] (N) {\texttt{Nation}};
			\node[table, below=of N, dotted] (R) {\texttt{Region}};
			
			\path (C) edge[->,pil,thick] node {\scriptsize custkey} (O) ;
			\path (C) edge[pil,dotted] node {\scriptsize nationkey} (N) ;
			\path (N) edge[pil,dotted] node {\scriptsize regionkey} (R) ;
			\path (O) edge[->,pil,thick] node {\scriptsize orderkey} (L) ;
			\path (N) edge[pil,dotted] node {\scriptsize nationkey} (S) ;
			\path (L) edge[<-,pil,thick] node[text width=0.9cm] {\scriptsize partkey suppkey} (P) ;   
			\path (PS) edge[->,pil,thick] node {\scriptsize partkey} (P) ; 
			\path (P) edge[pil,dotted] node {\scriptsize suppkey} (S) ;
			\path (S) edge[->,pil,thick] node {\scriptsize suppkey} (L) ; 
			\path (L) edge[pil,thick,bend left=55] node {\scriptsize partkey} (PS) ; 
			\end{tikzpicture}}
		\caption{\small{SDWithRedundancy}} 
		\label{fig:sdwith}
	\end{subfigure}%
	\begin{subfigure}{0.5\textwidth}
		\centering
		\resizebox{0.88\linewidth}{!}{
			\begin{tikzpicture}[node distance=1.2cm, auto]
			
			\node[table, fill=red!20] (C) {\texttt{Customer}};
			\node[table, right=of C, fill=red!20] (O) {\texttt{Order}};
			\node[table, right=of O, fill=red!20] (L) {\texttt{Lineitem}};
			\node[table, below=of O, fill=yellow!20] (O2) {\texttt{Order}};
			\node[table, right=of L,fill=red!20] (P2) {\texttt{Part}};
			\node[table, below=of L, fill=yellow!30] (L2) {\texttt{Lineitem}};
			\node[table, below=of L2,fill=yellow!30] (P) {\texttt{PartSupp}};
			\node[table, below=of P2,fill=yellow!30] (PS) {\texttt{Part}};
			\node[table, right=of R, dotted] (S) {\texttt{Supplier}};
			\node[table, below=of C, dotted] (N) {\texttt{Nation}};
			\node[table, below=of N, dotted] (R) {\texttt{Region}};
			
			\path (C) edge[<-,pil,thick] node {\scriptsize custkey} (O) ;
			\path (O) edge[<-,pil,thick] node {\scriptsize orderkey} (L) ;
			\path (L) edge[<-,pil,thick] node {\scriptsize partkey} (P2) ;
			\path (L2) edge[->,pil,thick] node[text width=0.9cm] {\scriptsize partkey suppkey} (P) ; 
			\path (L2) edge[<-,pil,thick] node[text width=0.9cm] {\scriptsize orderkey} (O2) ;  
			\path (PS) edge[<-,pil,thick] node {\scriptsize partkey} (P) ; 
			\path (L2) edge[->,pil,thick,bend left=55] node[text width=0.6cm] {\scriptsize partkey} (PS) ; 
			\path (L2) edge[pil,dotted] node {\scriptsize Replicated} (L) ;
			\path (O2) edge[pil,dotted] node {\scriptsize Replicated} (O) ;
			\path (PS) edge[pil,dotted] node[right] {\scriptsize Replicated} (P2) ;		
			\end{tikzpicture}}
		\caption{\small{WD}} 
		\label{fig:wd}
	\end{subfigure}
	\caption{\small{Partitioning designs used for experimentation}}
	\label{fig:partdesigns}
\end{figure*}
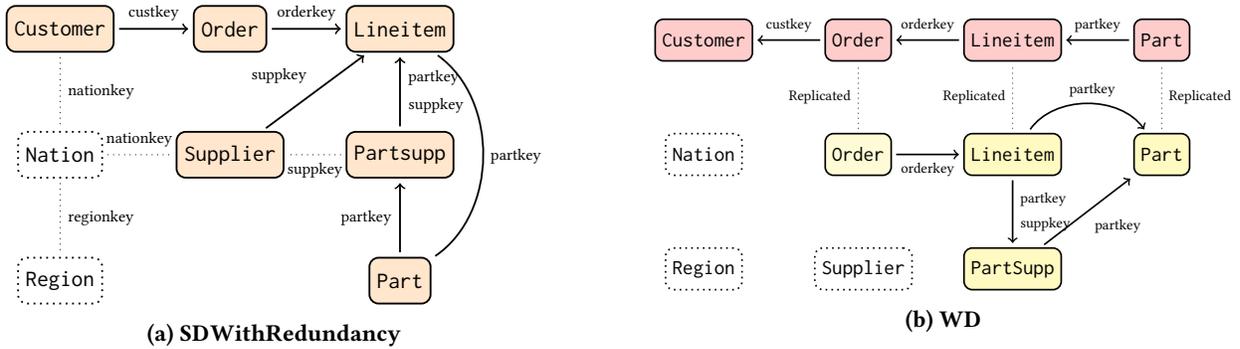

We now present empirical evidence of the accuracy of approximate answers obtained from a sample of the data that is co-hash partitioned. 

\introparagraph{Experimental Setup.} We run all experiments on System X, using a 100GB TPC-H data set, both uniform and
skewed. We generate the skewed data set using the tool from~\cite{tpchSkew} setting parameter $z=2$ (medium skew).  
Since our goal is not to study the already established performance benefits of co-hash partitioning, but to study its benefits for obtaining estimates with confidence bounds in failure/straggler scenarios, the system used is not  important.

\introparagraph{Partitioning Designs.} 
We experiment with co-hash partitioning designs validated and recommended by previous work~\cite{locationAware}. There are three designs we consider:
\begin{enumerate}
	\item SDWithoutRedundancy: This is the schema driven (SDWithout) partitioning design recommended in~\cite{locationAware} as the best co-hash partitioning strategy without redundancy for the TPC-H benchmark schema. This is depicted in Figure~\ref{fig:hierarchy}.
	\item SDWithRedundancy: This is the schema driven (SDWith) partitioning design suggested in~\cite{locationAware} as the best co-hash partitioning strategy with tuple-level redundancy. This is graphically presented in Figure~\ref{fig:sdwith}; the root of the co-hashing hierarchy is relation {\tt Lineitem} and all other tables are co-located through it.
	\item WD: This is the workload driven (WD) partitioning design suggested by the algorithm in~\cite{locationAware}. It is presented in Figure~\ref{fig:wd}. It has two hierarchies with relation level redundancy for tables {\tt Lineitem}, {\tt Orders} and {\tt Part} along with tuple level redundancy.
\end{enumerate}

The design choice not only determines the accuracy of the estimation for the queries, but also affects feasibility of approximation. Depending on the database partitioning design, it may or may not be possible to provide approximate answers to the query as we will see next.


\begin{table*}
\centering
\small
\begin{tabular}{|c|c|c|c|c|c|c|c|c|c|c|c|c|c|c|c|c|c|c|c|c|c|c|c|c|c|}
\hline
Q & 1 & 2 & 3 & 4 & 5 & 6 & 7 & 8 & 9 & 10 & 11 & 12
& 13 & 14 & 15 & 16 & 17 & 18 & 19 & 20 & 21 & 22 & Total Y\\
 \hline
$SDWithout$ & Y & N & Y & Y & Y & Y & Y & Y & Y & Y & N & Y 
& Y & Y & N & N & N & Y & Y & N & Y & N & 15\\
$SDWith$ & Y & N & Y & N & Y & Y & Y & Y & Y & Y & N & Y 
& N & Y & N & N & N & N & Y & N & N & N & 11 \\
$WD$ & Y & N & Y & N & Y & Y & Y & Y & Y & Y & N & Y 
& N & Y & N & N & Y & N & Y & N & N & N & 12 \\
\hline
\end{tabular}
\caption{{\small{Approximability of TPCH queries for the 3 designs considered.}}}
\label{tab:asapapprox}
\end{table*}

\introparagraph{Queries.} 
The TPC-H benchmark consists of 22 analytical queries, out of which 11 queries have nested predicates. Table~\ref{tab:asapapprox} lists the queries that are approximable for each of the three designs we consider; `Y' indicates that the query is approximable for that design and `N' indicates that is not. Notice that for SDWithout, 15 queries are approximable, and for design SDWith only 11 queries are approximable; queries 4, 13, 18, and 21 that have nested aggregates are approximable by SDWithout, but not by the other two designs. Query 17 is approximable by only WD. However, as we will see later, SDWith provides estimates with higher accuracy.


\subsection{Single Relation}
\begin{figure}
	\centering
	\includegraphics[scale=0.45]{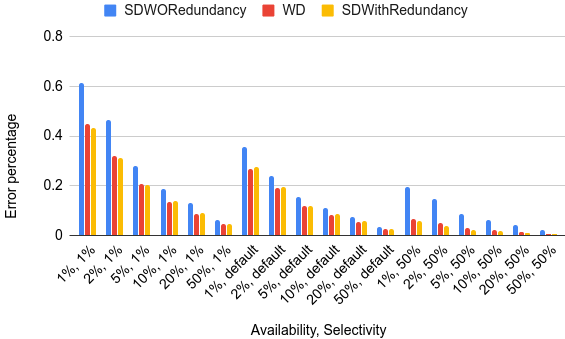}	
	\caption{Change in accuracy with availability and selectivity: single relation}
	\label{fig:query6_uniform}
\end{figure}

We first experiment with single table queries; we consider Query 6 from the TPC-H benchmark that involves a single table: {\tt Lineitem}. We change the availability of the data, i.e., we change the availability of data for the respective hierarchies for all three designs, and the selectivity of the predicate for the Lineitem relation. We present the results on accuracy for the uniform data set in Figure~\ref{fig:query6_uniform}.

The errors for all designs are very low and decreases with increase in availability and selectivity for all three designs. The accuracy of the SDWith design is slight better than the res and this suggests that increasing
the length of the hierarchy exacerbates errors; hierarchy length
for SDWith is 1, whereas the hierarchy length for SDWithout and WD is 3 and 2, respectively.

For the skewed data set (not presented due to space constraints),
this does not hold. Designs SDWith and SDWithout provide similar results, with slightly higher error rates (below 5\%). However, design WD results in very high errors due to high skew in the number and values of lineitems
associated with each part element. This suggests that apart from
cluster sizes and length of hierarchy, skew needs to be taken into
account for partitioning designs.

\begin{figure*}
	\begin{subfigure}{0.5\textwidth}
		\centering
		\includegraphics[scale=0.45]{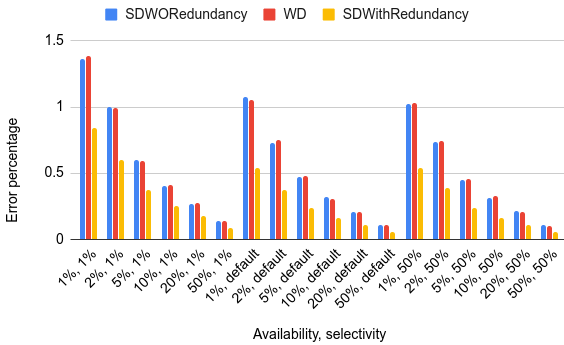}
		\caption{Uniform data}
		\label{fig:query3_uniform}
	\end{subfigure}%
	\begin{subfigure}{0.5\textwidth}
		\centering
		\includegraphics[scale=0.45]{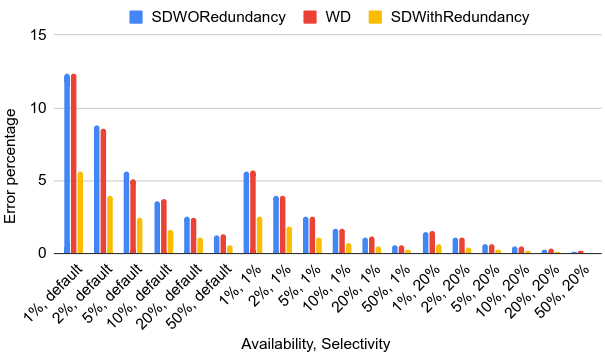}
		\caption{Skewed data}
		\label{fig:query3_skew}
	\end{subfigure}%
	\caption{Change in accuracy with availability and selectivity of a single hierarchy}
\end{figure*}

\subsection{A single hierarchy}
Next, we look at Query 3 of the benchmark that involves the join of {\tt Customer}, {\tt Orders} and {\tt Lineitem}. We present results for the uniform and skewed data set in Figures~\ref{fig:query3_uniform} and~\ref{fig:query3_skew}, respectively.
For both data-sets, the accuracy of all three designs improves with availability and selectivity. Although, the errors for the skewed data-set are much higher compared to the uniform data-set, at higher availability (and also selectivity), the errors come within 2\%. Even when 80\% of the data is unavailable, we obtain results within 5\% accuracy.

\subsection{Multiple hierarchies}
\begin{figure}
	\centering
		\includegraphics[scale=0.4]{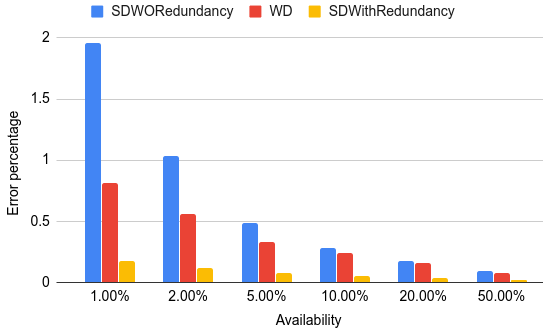}
	\caption{Change in accuracy with availability for multiple hierarchies}
		\label{fig:query9_uniform}
\end{figure}

We now consider Query 9 of the benchmark that involves join on {\tt Orders}, {\tt Lineitem}, {\tt Part}, {\tt PartSupp}, {\tt Supplier} and {\tt Nation}. This involves data across two hierarchies for SDWithout and WD, but a single hierarchy for SDWith. We change the availability of all the hierarchies involved and plot the accuracies obtained in Figure~\ref{fig:query9_uniform} for the uniform data set. We observe that even at 1\% availability, all three designs are accurate within 2\% error. However, with the skewed data set, SDWith provides similar errors, but SDWithout and WD produce very high errors due to the skewed number of tuples co-located for each root tuple. 

\subsection{Highly selective query}
\begin{figure}   
		\includegraphics[scale=0.4]{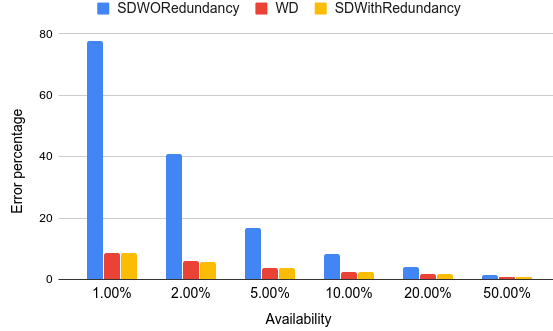}
	\caption{Change in accuracy with availability for a highly selective query}
	\label{fig:highly_selective}
\end{figure}

Query 19 of the TPC-H benchmark is a highly selective query; it involves the join of {\tt Lineitem} and {\tt Part} and the result only contains 0.001\% of the {\tt Lineitem} table. For SDWith and WD designs, this involves a single hierarchy, whereas for the SDWithout design, it involves two hierarchies. We vary the availability of all hierarchies and plot the result in Figure~\ref{fig:highly_selective}. We observe that while SDWith and WD provide estimates with higher accuracy than SDWithout for lower availabilities. At 20\% availability, all designs provide estimates within 5\% error. For the skewed data set, SDWith provides errors within 5\% whereas estimates from the other two designs result in high errors since most samples do not satisfy the predicates.

\subsection{Query with nested predicates}
\begin{figure}
	\centering
		\includegraphics[scale=0.4]{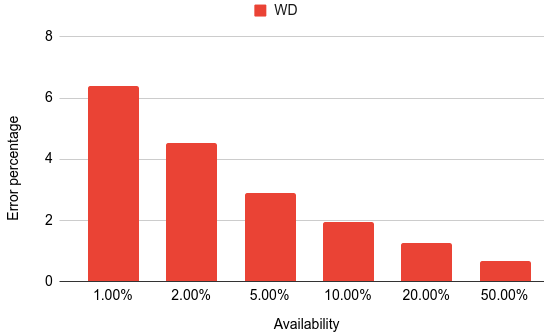}
	\caption{Change in accuracy with availability for a query with nested predicates}
		\label{fig:nested_predicate}
\end{figure}

We consider Query 17 of the benchmark which involves a nested predicate on {\tt Lineitem}. As shown in Table~\ref{tab:asapapprox}, the query is approximable only the WD design and not the other two designs. The change in accuracy with availability is plotted in Figure~\ref{fig:nested_predicate}.

\subsection{Queries with Group By}
\label{subsec:groupby}
We now discuss obtaining approximate results for queries with Group By.
We consider queries 3 and 9 along with the {\tt Group By} operation. We measure the change in percentage of missing groups with availability for different designs. The accuracy of estimated aggregate values are similar to the results obtained without Group By.

\introparagraph{Query 9.} The true result has a total of 175 groups
and the grouping attributes involve {\tt n\_name} and {\tt
  o\_orderdate}. For the uniform data set, all groups are present in the result at 1\% availability. For the skewed data set, however, we need 80\% availability to obtain about 80\% of the groups and 95\% availability to obtain most of the groups. (SDWith performs better than SDWithout and WD and obtains all groups by about 85\% availability.)

\introparagraph{Query 3.} The grouping attributes in this query are {\tt
  l\_orderkey}, {\tt o\_orderdate,} and {\tt o\_shippriority} and the query requests for the top 10 results. At about 80\% availability, SDWithout provides 8 out of 10 results and about 90\% availability provides all 10 results in a consistent manner. For SDWith and WD, we need a high availability of 95\% to get about 8 matches. For the skewed data set, this result only gets exacerbated. This is primarily because the gap in aggregated values between groups is not high and hence even a small error will lead to the group being eliminated from the set of top 10.

%

\subsection{Summary of experimental results}

Our experimental results show that, even at very low-availability we can obtain estimates with high accuracy for aggregate queries by exploiting the statistical properties of co-hash partitioned databases. We also found that the errors can increase with the size of the hierarchy, the number of unavailable hierarchies, and with high skew. But, for the case of failures and stragglers, where we expect a reasonable (at least 20\%) amount of the data to be available, co-hash partitioning can be exploited to obtain approximate answers instead of errors.

%% file: conc.tex
\section{Conclusions}
\label{sec:conc}
In this paper we proposed a novel idea to exploit an efficient partitioning
strategy used in distributed systems, \emph{co-hash partitioning}, to address failures and stragglers that affect data analytic users. The key idea was to determine the statistical properties of available data in the event of a failure/straggler, and to use that to provide approximate answers with error bounds.
We validated our idea through extensive experiments on the TPC-H benchmark; in particular we found that accuracy of approximate answers for queries increases when the join paths and group-by attributes match the hierarchies embedded in the co-hash partitioning. 
One particularly promising avenue for future research is to exploit the same idea to provide approximate answers in the non-failure case; in the case where users choose to obtain approximate query processing.
Applying co-hash partitioning over factorization trees~\cite{fdbEngine, fdb} to provide quick approximations is another interesting area for future work.